\documentclass[pra,aps,twocolumn]{revtex4}
\usepackage[dvips]{graphicx}
\usepackage{amsmath}

\begin{document}

\title{Teleportation with insurance of an entangled atomic state via 
cavity decay}

\author{Grzegorz Chimczak, Ryszard Tana\'s and Adam Miranowicz}

\affiliation{Nonlinear Optics Division, Physics Institute, Adam
 Mickiewicz University, 61-614 Pozna\'n, Poland}

\date{\today}
\email{chimczak@kielich.amu.edu.pl}


\begin{abstract}
We propose a scheme to teleport an entangled state of two
$\Lambda$-type three-level atoms
via photons. The teleportation protocol involves the local
redundant encoding protecting the initial entangled state and
allowing for repeating the detection until quantum information
transfer is successful.
We also show how to manipulate a state of many $\Lambda$-type
atoms trapped in a cavity.
\end{abstract}

\maketitle

\thispagestyle{empty}

\section{Introduction}
Practical quantum computation requires considering systems containing
scalable number of qubits. Recently, schemes have been proposed
that employ more than two qubits to perform various quantum information
tasks~\cite{pell95:_decoh,duan:_effic,miranowicz:_gener}. There is also an interest in 
performing quantum teleportation of state of more than one qubit. 
Lee~\cite{lee01:tel_ent} has presented a setup for teleportation 
of an entangled state of two photons.
The scheme, as some other schemes~\cite{duan_nature,cabrillo99,bose,
browne_entanglement,feng_entanglement,duan:_effic,simon_entanglement,
clark_entanglement,zou_4distant}, uses photons because they propagate
fast and can carry quantum information over long distances.
On the other hand, photonic states are much worse for the storage of
the quantum information than atomic states. The scheme does not
provide a way to store the quantum information and therefore
it will be difficult to use in quantum computing.
Other problem is that the scheme works only with a $50\%$ success rate.
Bose \emph{et al.}~\cite{bose} have proposed a novel scheme to teleport state
of one atom using photonic states as a carriers of quantum
information. In the scheme, the quantum information is stored in atomic states but 
the probability of successful teleportation is about $49\%$.
The protocol of the teleportation can be repeated to teleport an
entangled state of two atoms. This method, however, has at the most
only a $25\%$ success rate.

In the present work, we propose a scheme that allows the teleportation
of an entangled state of two atoms with insurance. Our device employs
an atomic states for storage and photonic states to transfer quantum
information. There are two distinguishing features of our protocol.
First of them is that the probability of successful teleportation of the initial
entangled state is about $49\%$. Second of them is that the initial
state is not lost, when detection stage is unsuccessful, because of
using \emph{local redundant encoding}~\cite{Enk97}. Hence the teleportation 
procedure can be repeated until the quantum transfer is successful.

The paper is organized as follows: In Sec. II, we describe the
teleportation device in detail.
In Sec. III, we show the operations manipulating a state of many
atoms trapped in a cavity.
In Sec. IV, we present the protocol of the teleportation. 
Sec. V gives the numerical results.

\section{Model }
\begin{figure}[htbp]
\includegraphics[width=8cm]{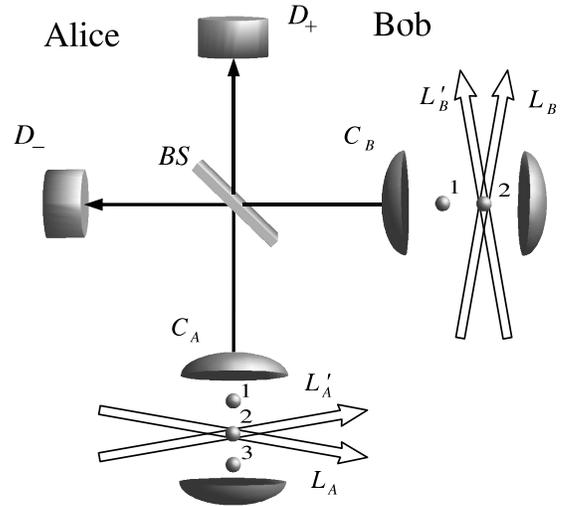}
\caption{Schematic
representation of the entangled-state teleportation device. The state of 
Alice's atoms $1$ and $2$ is teleported to Bob's atoms $1$ and $2$.}
\label{fig:rys1.eps}
\end{figure}

We consider the device composed of one cavity with three atoms
inside, one cavity with two atoms inside, a 50-50 beam splitter, two
lasers $L_{A}$ and $L_{B}$ with right- and two lasers 
$L_{A}^{\prime}$ and $L_{B}^{\prime}$ with left-circular polarized radiation  and 
two detectors $D_{+}$ and $D_{-}$. The system is shown in Fig.~\ref{fig:rys1.eps}.
The atoms are assumed to be located in fixed positions along a line in a linear
trap or an optical lattice inside an optical cavity.
We also assume that the atoms are separated by at least one optical
wavelength so they can be addressed individually by two different
laser fields. The propagation directions of the two laser beams are
very close to each other as to allow for effective transfer of photons from one beam
to the other mediated by the atom. Introducing two laser beams allows
for reseting the atomic states.

\begin{figure}[htbp]
  \begin{center}
\includegraphics[width=7cm]{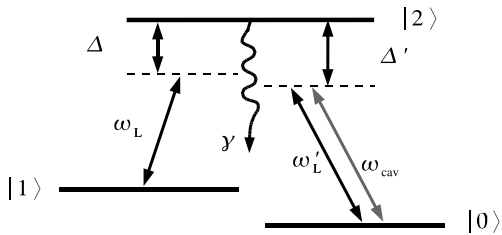}
\caption{Level scheme of one
of the identical $\Lambda$ atoms interacting with two classical laser fields
and the quantized cavity mode.} \label{fig:rys2.eps}
  \end{center}
\end{figure}

The cavity with two atoms inside and two lasers ($L_{B}$, $L_{B}^{\prime}$)
with different polarizations  belong to Bob. The sender -
Alice has the other parts of the device. All the trapped atoms are
modeled by three-level $\Lambda$ systems with an excited state
$|2\rangle$ and two ground states $|0\rangle$ and $|1\rangle$
as shown in Fig.~\ref{fig:rys2.eps}. The
excited state spontaneously decays with a rate $\gamma$. The
transition $|0\rangle \leftrightarrow |2\rangle$ is coupled to the cavity mode
with frequency $\omega_{\text{cav}}$ and coupling strength $g$. The
transition is also driven by a classical laser field with
frequency $\omega_{\text{L}}^{\prime}$ which is the same as the cavity mode
frequency. The coupling strength for this transition is denoted by
$\Omega^{\prime}$. Another classical laser field with different
polarization couples to the $|1\rangle \leftrightarrow |2\rangle$ transition with
the coupling strength $\Omega$. The frequency of the laser field
is $\omega_{\text{L}}$. We define two detunings
$\Delta=(E_{2}-E_{1})/\hbar-\omega_{\text{L}}$ and
$\Delta^{\prime}=(E_{2}-E_{0})/\hbar-\omega_{\text{cav}}$. 

The evolution of the system is conditional. Photon
detection corresponds to action of the operator
\begin{eqnarray}
  \label{eq:operatorColl}
  C&=&\sqrt{\kappa} (a_{A}+\epsilon a_{B}) \, ,
\end{eqnarray}
where $a_{A}$ and $a_{B}$ denote the annihilation operators for Alice's and
Bob's cavity modes, respectively, $\kappa$ denotes the cavity decay rate and
$\epsilon$ is equal to unity when there is a
click in the detector $D_{+}$ or minus unity for a click in $D_{-}$.
Between the emissions evolution of the system is governed by the
effective non-Hermitian Hamiltonian ($\hbar=1$)
\begin{eqnarray}
  \label{eq:Hamiltonian0}
  H&=&\sum_{k} (\Delta - i \gamma) \sigma_{22}^{(k)} - \sum_{k}
  \Delta_{r}  \sigma_{00}^{(k)}-i \kappa
  a^{\dagger} a \nonumber \\
   &&+\sum_{k}(\Omega \sigma_{21}^{(k)}+g a \sigma_{20}^{(k)}+\Omega^{\prime} \sigma_{20}^{(k)}+ {\rm{H.c.}}) \, ,
\end{eqnarray}
where $\Delta_{r}=\Delta^{\prime}-\Delta$.
In~(\ref{eq:Hamiltonian0}) we define $\sigma_{ij}^{(k)}\equiv
(|i\rangle \langle j|)_{k}$,
where $i,j=0,1,2$ for the $k$th atom. In the far off resonance
limit when $\Delta \gg \Omega$ and $\Delta^{\prime} \gg
\Omega^{\prime},g$, we can eliminate adiabatically the level $|2\rangle$
~\cite{pell,alexanian95,carmichaelKsiazkaMethods}. The conditions
have to be even more restrictive in our teleportation protocol because only
then we can properly estimate a phase shift factors for long evolution
times. Therefore we assume  ${10}^{-1} \Delta \gg \Omega$ and
${10}^{-1} \Delta^{\prime} \gg \Omega^{\prime},g$. In order to
simplify the
Hamiltonian~(\ref{eq:Hamiltonian0}) we also assume that $\gamma \ll
\Delta,\Delta^{\prime}$ and the product of the
excited level saturation parameter and the spontaneous decay rate is
much smaller than the decay rate of the cavity mode
($\gamma g^{2}/{\Delta^{\prime}}^{2},
\gamma {\Omega^{\prime}}^{2}/{\Delta^{\prime}}^{2},
\gamma {\Omega}^{2}/{\Delta}^{2} \ll \kappa$). Under these
conditions we can neglect the influence of the spontaneous decay rate
on teleportation. Otherwise, the probability of success will be much lower
as it was proved in~\cite{chimczak02:_effect}. These assumptions were
also used in another quantum information process of entangled state
preparation~\cite{beige:_entan}. With these assumptions, after
adiabatic elimination of the excited state $|2\rangle$ of the atoms, the
Hamiltonian takes the form
\begin{eqnarray}
\label{eq:Hamil1}
H&=&-\sum_{k} \Delta_{r} \sigma_{00}^{(k)} -i \kappa a^{\dagger} a \nonumber \\
 & &-\sum_{k}( \delta_{1} \sigma_{11}^{(k)}+\delta_{2}
 \sigma_{00}^{(k)}+\delta_{3} 
 a^{\dagger} a \sigma_{00}^{(k)}) \nonumber \\
 & &-\sum_{k}(\delta_{4} \sigma_{10}^{(k)} +\delta_{5} a
 \sigma_{10}^{(k)} +\delta_{6} a 
 \sigma_{00}^{(k)} +{\rm{H.c.}})\, ,
\end{eqnarray}
where $\delta_{1}=\Omega^{2} / \Delta$, $\delta_{2}=
{\Omega^{\prime}}^{2}/\Delta^{\prime}$, $\delta_{3}=g^{2}/\Delta^{\prime}$,
$\delta_{4}= \Omega\Omega^{\prime}(\Delta^{-1}+{\Delta^{\prime}}^{-1})/2$, 
$\delta_{5}=g\Omega (\Delta^{-1}+{\Delta^{\prime}}^{-1})/2$
and $\delta_{6}=g\Omega^{\prime}/ \Delta^{\prime}$.
The parameters $\delta_{1}-\delta_{6}$ account for various
contributions to the effective Hamiltonian, for example, $\delta_{4}$
describes the transfer of photon from one laser beam to the other via
coupling to the atom, $\delta_{6}$ describes the transfer of photon
from a cavity into the laser $L'$ beam, etc. In this approximations
all the atomic dynamics are restricted to the ground states
$|0\rangle$ and $|1\rangle$ which can be treated as atomic qubits.

\section{Quantum operations}

In our teleportation protocol we need certain transformations or quantum
operations, which applied to a given state of the system (quantum
register) transform it into another state. Such operations are
performed with the unitary evolution operator 
$e^{-iHt}$ applied to the state of the system. It is assumed that only
one atom is illuminated at a time and that the laser fields are such
that $\Omega \gg \Omega^{\prime} \gg g$. It is useful to distinguish
between the results of the action of the evolution operator onto
particular states of the system and write down explicitly the results
for some special cases. We list a number of local operations that can
be performed by Alice and Bob on their states.

To fix and simplify the notation we label the Alice atoms with numbers $(1,2,3)$ and
Bob's atoms with numbers $(1,2)$.
Let us denote a state of the system of $n=\text{(2 or 3)}$ atoms
trapped in the cavity with $y$ 
photons by $|x_{1} \dots x_{n} y\rangle$ where $x_{k}$ is the $k$th
atom state ($k=1, \dots ,n$). This means that the state of the Alice
part has the form $|x_{1}x_{2}x_{3}y\rangle$ and the Bob part has the
form $|x_{1}x_{2}y\rangle$, for example,
$|1110\rangle=|1\rangle_{1}|1\rangle_{2}|1\rangle_{3}|0\rangle$ means that
at the Alice site atoms $(1,2,3)$ are all in their state $|1\rangle$
and the cavity field is in state $|0\rangle$. In our notation, the first
two or three numbers in the ket denote atomic states with labels
increasing from left to right, and the last number is
reserved for the field state.  The joint state of the entire system
can be described in the basis that is formed by the product states of
the Alice part and the Bob part. Later on, we use the simplified
notation
$|x_{1}x_{2}x_{3}y\rangle_{A}\otimes|x_{1}x_{2}y\rangle_{B}=|x_{1}x_{2}x_{3}y;x_{1}x_{2}y\rangle$.

The simplest operation is just waiting for an arbitrary time $t$ while
all of the lasers are turned off. In this case both $\Omega $ and
$\Omega^{\prime}$ are set to zero, and we can use the simplified
Hamiltonian given by
\begin{equation}
\label{eq:Hamil_Bez}
H=-\sum_{k=1}^{n} (\Delta_{r} \sigma_{00}^{(k)}+\delta_{3} a^{\dagger} a
\sigma_{00}^{(k)}) -i \kappa a^{\dagger} a \, .
\end{equation}
 During this operation the evolution of the system is given by 
\begin{equation}
\label{eq:Oper_0}
e^{-iHt} |x_{1} \dots y\rangle=
e^{i N_{0} (\Delta_{r}+y \delta_{3}) t} e^{-y \kappa t}|x_{1} \dots y\rangle \, ,
\end{equation}
where $N_{0}$ is the number of atoms being in state $|0\rangle$ that
are not illuminated by the laser field.
In order to simplify the following transformations we assume 
$\delta_{1}=\Delta_{r}$.
Moreover, we want the probability of no collapse during the encoding stage to be
close to unity. This can be done provided that $\delta_{5} \gg \kappa$.
Those assumptions imply that $\Delta_{r} \gg \kappa , \delta_{3}$ and therefore
the Hamiltonian can be written as $H=-\Delta_{r} \sigma_{00}^{(k)}$ for
the operation time $t \approx \Delta_{r}^{-1}$. Thus, for short times the
evolution simplifies to
\begin{equation}
\label{eq:Oper_0'}
e^{-iHt} |x_{1} \dots y\rangle=
\alpha^{N_{0}}(t) |x_{1} \dots y\rangle \, ,
\end{equation}
where $\alpha(t)=e^{i \Delta_{r} t}$.

As the next local operation we consider the illumination of the $k$th
atom by the laser field driving $|1\rangle \leftrightarrow |2\rangle$
transition ($\Omega\ne 0$) while the
second laser field coupled to
$|2\rangle \leftrightarrow |0\rangle$ is turned off
($\Omega^{\prime}=0$). We can use this
laser field to get a number of useful transformations. The transformations times
are of the order of $\delta_{5}^{-1}$ and therefore under conditions
$\delta_{1}=\Delta_{r}$ and $x_{k}+y>0$ the system evolution can be
well approximated by the relation
\begin{eqnarray}
\label{eq:Oper_inne}
e^{-iHt} |x_{1} \dots x_{k} \dots y\rangle&=&
f_{N_{0}, \beta_{k}}(t)
(\cos( \xi_{k} t) |x_{1} \dots x_{k} \dots y\rangle  \nonumber \\
&&+ i \sin( \xi_{k} t)|x_{1} \dots x_{k}^{\prime} \dots y^{\prime}\rangle ) \,
\end{eqnarray}
where
\mbox{$\xi_{k}=\sqrt{\beta_{k}} \delta_{5}$},
\mbox{$\beta_{k}=x_{k}+y$},
$f_{N_{0}, \beta_{k}}(t)=\alpha^{N_{0}+1}(t)
\exp\left\{i\, \delta_{3} [\beta_{k}(2 N_{0}+1)-N_{0}] t /2\right\}$,
$x_{k}^{\prime}=x_{k}-(-1)^{x_{k}+1}$
and $y^{\prime}=y+(-1)^{x_{k}+1}$.
One can see that we are able to perform different transformations by
illuminating the $k$th atom for different times.

We map the atomic state onto the cavity mode by choosing the interaction time
$t^{(1)} =(\pi / 2 + 2 n \pi)/ \delta_{5}$ where $n$ is an integer,
according to
\begin{eqnarray}
\label{eq:UD}
|x_{1} \dots 1 \dots 0\rangle & \rightarrow & i f_{N_{0}, 1}(t^{(1)})
 |x_{1} \dots 0 \dots 1\rangle \, , \\
\label{eq:UD_b}
|x_{1} \dots 0 \dots 1\rangle & \rightarrow & i f_{N_{0}, 1}(t^{(1)})
 |x_{1} \dots 1 \dots 0\rangle \, .
\end{eqnarray}

If we turn the laser on for time $t^{(2)}=(\pi/4 + 2 n \pi)/
\delta_{5}$ then we create a maximally entangled 
state of the illuminated atom and the cavity system according to
\begin{eqnarray}
\label{eq:ud}
|x_{1} \dots 1 \dots 0\rangle & \rightarrow & 
\frac{f_{N_{0}, 1}(t^{(2)})}{\sqrt{2}}
 (i |x_{1} \dots 0 \dots 1\rangle \nonumber \\
 &&+ |x_{1} \dots 1 \dots 0\rangle) \, ,\\
|x_{1} \dots 0 \dots 1\rangle & \rightarrow &
\frac{f_{N_{0}, 1}(t^{(2)})}{\sqrt{2}}
 (i |x_{1} \dots 1 \dots 0\rangle \nonumber \\
 &&+ |x_{1} \dots  0 \dots 1\rangle) \, .
\end{eqnarray}
Interaction for time $t^{(3)}=(2 n \pi)/ \delta_{5}$ gives only the phase factor
\begin{eqnarray}
\label{eq:N}
|x_{1} \dots x_{k} \dots y\rangle &\rightarrow &
f_{N_{0}, \beta_{k}}(t^{(3)})  |x_{1} \dots x_{k} \dots y\rangle \, .
\end{eqnarray}

The transformations can also be performed for the states with $\beta_{k}=2$.
The states mapping can be done for the illumination time
$t^{(4)} =(\pi/2 +2 m \pi)/ (\sqrt{2} \delta_{5})$ where $m$ is an integer as
\begin{eqnarray}
\label{eq:UD2}
|x_{1} \dots 1 \dots 1\rangle & \rightarrow &
i f_{N_{0}, 2}(t^{(4)}) |x_{1} \dots 0 \dots 2\rangle \, , \\
\label{eq:UD2_b}
|x_{1} \dots 0 \dots 2\rangle & \rightarrow &
i f_{N_{0}, 2}(t^{(4)}) |x_{1} \dots 1 \dots 1\rangle \, .
\end{eqnarray}
Interaction for the time $t^{(5)}=(\pi /4 +2 m \pi)/ (\sqrt{2} \delta_{5})$
leads to generation of maximally entangled state
\begin{eqnarray}
\label{eq:ud2}
|x_{1} \dots 1 \dots 1\rangle & \rightarrow & \frac{1}{\sqrt{2}}
  f_{N_{0}, 2}(t^{(5)})
  (i |x_{1} \dots 0 \dots 2\rangle \nonumber \\
  &&+ |x_{1} \dots 1 \dots 1\rangle) \nonumber \, ,\\
|x_{1} \dots 0 \dots 2\rangle & \rightarrow & \frac{1}{\sqrt{2}}
  f_{N_{0}, 2}(t^{(5)})
  (i |x_{1} \dots 1 \dots 1\rangle \nonumber \\
  &&+ |x_{1} \dots 0 \dots 2\rangle)
\end{eqnarray}
and illumination for the time $t^{(6)} =(2 m \pi)/ (\sqrt{2} \delta_{5})$
generates only a phase factor.

There is a special state with $\beta_{k}=0$. If laser is turned on, the state
accumulates a phase shift but the population of the state remains unchanged
as described by
\begin{eqnarray}
\label{eq:beta0}
e^{-iHt} |x_{1} \dots 0 \dots 0\rangle &=&
\alpha^{N_{0}+1}(t) |x_{1} \dots 0 \dots 0\rangle \, .
\end{eqnarray}
When we want to map an arbitrary superposition of the two atomic
ground states onto the cavity mode then this feature of the state is very 
desired. However, when the detection stage is unsuccessful in our protocol,
the population transfer is necessary to repeat the teleportation process.
Therefore we have to use another local
operation consisting in simultaneous applying of two laser fields with different
polarizations (for instance $L_{A}$ and $L_{A}^{\prime}$). For evolution times
of the order of $t \approx \delta_{4}^{-1}$
we can neglect in the Hamiltonian all terms much smaller than $\delta_{4}$, then
the approximate state dynamics are given by
\begin{eqnarray}
\label{eq:Oper_powrotOgolnie}
e^{-iHt} |x_{1} \dots 0 \dots 0\rangle &=& \alpha^{N_{0}+1}(t)
(i \sin(\delta_{4} t)|x_{1} \dots 1 \dots 0\rangle \nonumber \\
&&+\cos( \delta_{4} t) |x_{1} \dots 0 \dots 0\rangle ) \, , \\
\label{eq:Oper_powrotOgolnieB}
e^{-iHt} |x_{1} \dots 1 \dots 0\rangle &=& \alpha^{N_{0}+1}(t)
(i \sin(\delta_{4} t)|x_{1} \dots 0 \dots 0\rangle \nonumber \\
&&+\cos( \delta_{4} t) |x_{1} \dots 1 \dots 0\rangle ) \, .
\end{eqnarray}
It is evident that by using $\pi /2$ pulse we can change the atom state even
if the cavity field mode is empty. This case can be described by
\begin{eqnarray}
\label{eq:Oper_powrot}
|x_{1} \dots 0 \dots 0\rangle &\rightarrow &
i \alpha^{N_{0}+1}(t^{(7)}) |x_{1} \dots 1 \dots 0\rangle \, , \\
\label{eq:Oper_powrotB}
|x_{1} \dots 1 \dots 0\rangle &\rightarrow &
i \alpha^{N_{0}+1}(t^{(7)}) |x_{1} \dots 0 \dots 0\rangle \, ,
\end{eqnarray}
where $t^{(7)}=\pi/(2 \delta_{4})$.

The set of transformations listed above forms necessary ingredients
for the teleportation protocol we present in the next Section.

\section{The teleportation protocol}
The teleportation protocol we propose here makes it possible to
teleport an entangled state of two atoms with insurance.
The state to be teleported is an entangled state of atoms $1$ and $2$
at Alice site, which is
\begin{equation}
  \label{eq:statezero}
|\psi_{0}\rangle=a\, |1\rangle_{1}
|0\rangle_{2} + b\, |0\rangle_{1} |1\rangle_{2}.
\end{equation}
The third atom in the Alice cavity and both Bob's atoms are prepared in
their states $|1\rangle$. The field modes in both cavities are initially
empty. Thus, the joint state of the entire system is initially given by
\begin{eqnarray}
\label{eq:Psi0}
|\Psi(0)\rangle &=&(a\, |1010\rangle_{A} + b\, |0110\rangle_{A} ) \otimes
 |110\rangle_{B}\nonumber\\
&=&a\,|1010;110\rangle + b\,|0110;110\rangle\, .
\end{eqnarray}
We assume that this state is given or prepared before the protocol starts.

The teleportation protocol consist of five stages: (i) preparation
stage, (ii) encoding stage, (iii) detection stage I, (iv) detection
stage II and (v) recovery stage.
In each stage there are a number of steps to be performed in order to
get, finally, the required result.

\subsection*{(i) preparation stage}
The aim of the preparation stage of the protocol is to create a
maximally entangled state of the third Alice atom and the first Bob atom.
This can be done by following the distant atom entangling technique of
Ref.~\cite{chimczak:_entanglement}. This stage consists of three steps.

\emph{(a)} First, Alice and Bob perform
transformation given by~(\ref{eq:UD}). They simply illuminate,
using lasers $L_{A}$ and $L_{B}$ their
atoms, {\em i.e.}, Alice's atom 3 and Bob's atom 1,
for the time $t_{1}=\pi /(2 \delta_{5})$.
After this operation each cavity is in one photon state.

\emph{(b)} Next, they wait until either of Alice's detectors clicks.
All lasers are turned off and therefore, before the detection event,
evolution of their systems is described by~(\ref{eq:Oper_0'}).
One photon registered by Alice corresponds to an action of the collapse
operator~(\ref{eq:operatorColl}) and leads to creation of maximally
entangled state of both cavity fields.

\emph{(c)} After the detection event, Alice and Bob have to turn on the lasers
$L_{A}$ and $L_{B}$, immediately.
They illuminate the two atoms for time $t_{1}$ performing the transformation
given by~(\ref{eq:UD_b}). This operation leads to mapping and storage
of the entangled state of both cavity fields in the state of
Alice's atom 3 and Bob's atom 1.
This concludes creating a maximally entangled state of the two atoms
and then the global system state is given by
\begin{eqnarray}
\label{eq:Psi1}
|\Psi\rangle &=& a |1000;110\rangle + a \epsilon e^{\frac{i}{2} \delta_{3}
  t_{1}} |1010;010\rangle \nonumber \\
&&+ b |0100;110\rangle + b \epsilon e^{\frac{i}{2} \delta_{3} t_{1}}
  |0110;010\rangle \, .
\end{eqnarray}

\subsection*{(ii) encoding stage}
The encoding stage is introduced to apply the local redundant
encoding~\cite{Enk97} in which Alice codes the entangled state of her
first two atoms (atoms 1 and 2), that is to be teleported,
to the entangled state of four atoms (atoms 1,2,3 of Alice and atom 1
of Bob). The third Alice atom and the first Bob atom are the backup
atoms which allow to protect the teleported
state in case of protocol failure in the detection stage. The encoding consists
of a sequence of four steps.

\emph{(a)} First of them is mapping the state of the first Alice
atom onto the cavity mode by illuminating (using laser $L_{A}$)
the atom for time $t_{1}$. This corresponds to the
transformations given by~(\ref{eq:UD}) and~(\ref{eq:beta0}).
During the operation Bob's lasers are turned off and therefore he uses
transformation~(\ref{eq:Oper_0'}).
After the operation the unnormalized joint state becomes
\begin{eqnarray}
\label{eq:Psi2}
|\widetilde{\Psi}\rangle  &=&
i a \alpha_{1} e^{i \frac{3}{2} \delta_{3} t_{1}} |0001;110\rangle
+ i a \epsilon \alpha_{1} e^{i \frac{3}{2} \delta_{3} t_{1}}
|0011;010\rangle \nonumber \\
&&+ b |0100;110\rangle
+ b \epsilon e^{\frac{i}{2} \delta_{3} t_{1}} |0110;010\rangle  \, .
\end{eqnarray}
where $\alpha_{1}=\alpha(t_{1})$.

\emph{(b)} The second step of the encoding stage is illuminating the third Alice
atom. One can see that $\beta_{2}=1$ in all terms of the
superposition~(\ref{eq:Psi2}).
The purpose of the second operation is to make $\beta_{2}$'s different.
The Rabi frequency scales with $\beta$ and therefore we can perform
independly different transformations for different values of $\beta$.
Alice switches the laser $L_{A}$ on for the appropriate interaction time
leading to the transformations~(\ref{eq:UD}), (\ref{eq:UD_b}), (\ref{eq:UD2})
and~(\ref{eq:beta0}). It is clear that the time has to satisfy conditions
$t_{2} \delta_{5}=\pi/2 + 2 n \pi$ and $t_{2} \sqrt{2}
\delta_{5}=\pi/2 +2 m \pi$. This can be
done only approximately for $n=7$ and $m=10$.
During this step Bob waits with lasers turned off thus
evolution of state of his system is given by~(\ref{eq:Oper_0'}).
After this operation we achieve the state close to
\begin{eqnarray}
\label{eq:Fi3}
|\widetilde{\Psi}\rangle &=&
b |0100;110\rangle
+ i b \epsilon \alpha_{2} e^{i \delta_{3} (\frac{1}{2} t_{1}+t_{2})}
|0101;010\rangle  \nonumber \\
&&-a \alpha_{1} \alpha_{2} e^{i \frac{3}{2} \delta_{3}(t_{1}+t_{2})}
|0010;110\rangle  \nonumber \\
&&- a \epsilon \alpha_{1} \alpha_{2}^{2} e^{i \delta_{3} (\frac{3}{2} t_{1}+4
  t_{2})} |0002;010\rangle \, ,
\end{eqnarray}
where $\alpha_{2}=\alpha(t_{2})$. We neglect the low populated
states in the superposition~(\ref{eq:Fi3})
but we include them as all other imperfections of the operation
in our numerical calculations.

\emph{(c)} The third step is the most important at the encoding stage.
The previous two steps are intended to prepare the third one,
which creates the entangled state of three atoms and cavity field.
In order to make the entangled state Alice has to swap one pair of the state
amplitudes without exchanging the second pair of amplitudes. Alice can do
that by turning on the $L_{A}$ laser and illuminating her second atom for the
time which leads to completing the
transformations~(\ref{eq:N}), (\ref{eq:UD2}), (\ref{eq:UD2_b}) and
(\ref{eq:beta0}).
Here, we meet the same problem as in the second step because the
illuminating time has to satisfy two conditions: $t_{3} \delta_{5}=2 n \pi$ and
$t_{3} \sqrt{2} \delta_{5}=\pi/2 +2 m \pi$. We can find an approximate solution
for $n=3$ and $m=4$. In this step Bob's lasers are turned off.
Just as in the previous step, we neglect the states
for which the population is close to zero and obtain
\begin{eqnarray}
\label{eq:Psi4}
|\widetilde{\Psi}\rangle &=&
a \alpha_{1} \alpha_{2} e^{i \frac{3}{2} \delta_{3}(t_{1}+t_{2})} |0010;110\rangle \nonumber \\
&&+ i a \epsilon \alpha_{1} \alpha_{2}^{2} \alpha_{3}^{2}
e^{i \delta_{3} (\frac{3}{2} t_{1}+4 t_{2}+ 4 t_{3})}  |0101;010\rangle \nonumber \\
&&- b \alpha_{3} e^{i \frac{3}{2} \delta_{3} t_{3}} |0100;110\rangle \nonumber \\
&&+ b \epsilon \alpha_{2} \alpha_{3}^{2} e^{i \delta_{3} (\frac{1}{2}
  t_{1}+t_{2}+4 t_{3})}|0002;010\rangle \, ,
\end{eqnarray}
where $\alpha_{3}=\alpha(t_{3})$. Although the entangling is already done,
one can see that the state~(\ref{eq:Psi4}) is not protected yet. For instance,
if the two-photon state is detected then the initial state of the first two
Alice atoms will be lost.

\emph{(d)} In order to change the state~(\ref{eq:Psi4}) into a protected state
Alice illuminates her third atom using the $L_{A}$ laser.
This is the fourth step of the encoding
stage. Alice needs to perform transformations~(\ref{eq:N}), (\ref{eq:UD2_b})
and (\ref{eq:beta0}), therefore the illumination time has to fulfill
the conditions $t_{4} \delta_{5}=2 n \pi$ and
$t_{4} \sqrt{2} \delta_{5}=\pi/2 +2 m \pi$.
It is obvious that the time is the same as for the previous operation
time and thus $\alpha_{3}=\alpha_{4}=\alpha(t_{4})$.
We again neglect low populated states.
During this step Bob performs two operations.
First, Bob waits for time $t_{4}-t_{1}/2$ with lasers turned off.
Next he creates a maximally entangled
state of his second atom and his cavity. For this purpose he
turns the laser $L_{B}$ on for time $t_{1}/2$ performing
the transformation given by~(\ref{eq:ud}).
Alice and Bob perform their actions in such a way that they end
the fourth step at the same time.
Then the global system is given by
\begin{eqnarray}
\label{eq:Psi5}
|\widetilde{\Psi}\rangle &=&
(a \alpha_{1} \alpha_{2} e^{i \frac{3}{2} \delta_{3} (t_{1}+t_{2})}
|0010\rangle_{A}
- b |0100\rangle_{A}) \nonumber \\
&&\otimes ( i |101\rangle_{B} + |110\rangle_{B} ) \nonumber \\
&&+( i a \epsilon \alpha_{1} \alpha_{2}^{2} \alpha_{3}^{2}
e^{i \delta_{3} (\frac{7}{4} t_{1}+4 t_{2}+\frac{7}{2} t_{3})}
|0101\rangle_{A}  \nonumber \\
&&+ i b \epsilon \alpha_{2} \alpha_{3}^{3}
e^{i \delta_{3} (\frac{3}{4} t_{1}+ t_{2}+\frac{13}{2} t_{3})}
|0011\rangle_{A})  \nonumber \\
&&\otimes ( i |001\rangle_{B} + |010\rangle_{B} ) \, .
\end{eqnarray}

This is the end of the encoding stage.
If we wanted to store the protected state we would map the cavity state to
the first Alice atom state. Then we would have the entangled state of
four atoms. However, we want the photonic state to be the 
state~(\ref{eq:Psi5}) because we use the cavity field for quantum 
information transfer.

\subsection*{(iii) detection stage I}
The third stage of the protocol is the first detection stage, in which
Alice just waits for time $t_{D} \gg \kappa^{-1}$ making
a measurement of the fields  leaking from the cavities.
The detection of one photon only leads to the quantum information transfer.
If Alice does not detect any photon or detects two photons the teleportation
process will be unsuccessful. However, even then, quantum information will
be safe owing to the local redundant encoding.
In the absence of any laser field
the evolution is given by~(\ref{eq:Oper_0}).
If Alice does not detect any photon in this stage the state evolves into
\begin{eqnarray}
\label{eq:Psi7}
|\Psi\rangle &=&
- a \alpha_{1} \alpha_{2} e^{i \frac{3}{2} \delta_{3} (t_{1}+t_{2})}
|0010;110\rangle \nonumber \\
&&+ b |0100;110\rangle   \, .
\end{eqnarray}
This is one of two unsuccessful cases. The initial state which
Alice wanted to teleport is modified by phase shift factors but it is not lost.
The modified initial state is stored in the second and third Alice atoms.
In order to repeat the whole protocol Alice has to reset her first atom.
She turns on both her lasers ($L_{A}$ and $L_{A}^{\prime}$) for the time
$t_{5}=\pi/(2 \delta_{4})$ performing transformation~(\ref{eq:Oper_powrot}).
During the reseting operation both Bob's lasers are turned off.

If the evolution given by~(\ref{eq:Oper_0}) is interrupted by a collapse at
time $t_{j} < t_{D}$ then the jump operator $C$ acts on the global system state.
After that the transformation~(\ref{eq:Oper_0}) continues changing the state.
If Alice registers the second click of either of her detectors the joint state
becomes
\begin{eqnarray}
\label{eq:Fi_2_Coll}
|\Psi\rangle&=&
a \alpha_{1} \alpha_{2} \alpha_{3}^{-1}
e^{i \delta_{3} (t_{1}+3 t_{2}-3 t_{3})} |0100;000\rangle \nonumber \\
&&+b |0010;000\rangle  \, .
\end{eqnarray}
It is evident that the Alice initial state is not destroyed also in the second
case when the step is unsuccessful. Before the protocol can be repeated
Alice has to prepare her first atom in the state $|1\rangle$ and Bob
has to prepare both of his atoms in the state $|1\rangle$ using
transformation~(\ref{eq:Oper_powrot}). They reset the atoms in two
steps. First, Alice and Bob turn on all their lasers 
($L_{A}$, $L_{A}^{\prime}$, $L_{B}$ and $L_{B}^{\prime}$) for the time
$t_{5}=\pi/(2 \delta_{4})$. Alice and Bob illuminate their first atoms.
Next, Bob illuminates for the time $t_{5}$, using both his
lasers, his second atom while Alice waits with lasers turned off.

If there is no second photon detection then the quantum information transfer
is done and the global system state is given by
\begin{eqnarray}
\label{eq:Psi_Sukces}
|\widetilde{\Psi}\rangle
&=& a \epsilon \alpha_{1} \alpha_{2}^{2} \alpha_{3}^{2}
e^{i \delta_{3} (\frac{7}{4} t_{1}+4 t_{2}+\frac{7}{2} t_{3}+t_{j})}
|0100;010\rangle \nonumber \\
&&- b \epsilon_{1} |0100;100\rangle
+ a \epsilon_{1} \alpha_{1} \alpha_{2}
e^{i \frac{3}{2} \delta_{3} (t_{1}+t_{2})} |0010;100\rangle \nonumber \\
&&+ b \epsilon \alpha_{2} \alpha_{3}^{3}
e^{i \delta_{3} (\frac{3}{4} t_{1}+t_{2}+\frac{13}{2} t_{3}+t_{j})}
|0010;010\rangle \, .
\end{eqnarray}
After this stage Alice shares the information about her initial state with Bob.
Now Alice's initial state can be send to Bob, but it is also possible
for Bob to send it back to Alice.
We will not consider the case when Bob sends back the state.

\subsection*{(iv)  detection stage II}
In the fourth stage of the protocol Alice measures state of
her third atom. During the stage Bob waits with lasers turned off.
This stage consists of two steps.

\emph{(a)} First, Alice turns the $L_{A}$ laser on and illuminates
the third atom performing transformations~(\ref{eq:UD}) and (\ref{eq:beta0}).

\emph{(b)} After this she turns the laser off and the evolution is given
by~(\ref{eq:Oper_0}) which leads to the joint state
\begin{eqnarray}
\label{eq:Psi_Detekcja2_tlumienie}
|\widetilde{\Psi}\rangle
&=& a \epsilon \alpha_{1} \alpha_{2}^{2} \alpha_{3}^{2}
e^{i \delta_{3} (\frac{7}{4} t_{1}+4 t_{2}+\frac{7}{2} t_{3}+t_{j})}
|0100;010\rangle \nonumber \\
&&- b \epsilon_{1} |0100;100\rangle
+ e^{i (\Delta_{r}+3 \delta_{3}) t_{D}} e^{-\kappa t_{D}} \nonumber \\
&&\times \big[
i b \epsilon \alpha_{1} \alpha_{2} \alpha_{3}^{3}
e^{i \delta_{3} (\frac{9}{4} t_{1}+t_{2}+\frac{13}{2} t_{3}+t_{j})}
|0001;010\rangle \nonumber \\
&&+i a \epsilon_{1} \alpha_{1}^{2} \alpha_{2}
e^{i \frac{3}{2} \delta_{3} (2 t_{1}+t_{2})}
|0001;100\rangle \big] \, .
\end{eqnarray}
Alice again waits for time $t_{D}$ making a measurement of the fields
leaking from the cavities. The detection of one photon corresponds to
an action of the jump operator $C$ on the global
state~(\ref{eq:Psi_Detekcja2_tlumienie}). In this case the state becomes
\begin{eqnarray}
\label{eq:Psi_Detekcja2_1klik}
|\Psi\rangle
&=&  a \epsilon \epsilon_{1} \alpha_{1} \alpha_{3}^{-3}
e^{i \delta_{3} (\frac{3}{4} t_{1}+\frac{1}{2} t_{2}-\frac{13}{2} t_{3}-t_{j})}
|0000;100\rangle \nonumber \\
&&+  b |0000;010\rangle \, .
\end{eqnarray}
Otherwise, when Alice has not detected any photon, the joint state
is given by
\begin{eqnarray}
\label{eq:Psi_Detekcja2_0klik}
|\Psi\rangle &=&
- a \epsilon \epsilon_{1} \alpha_{1} \alpha_{2}^{2} \alpha_{3}^{2}
e^{i \delta_{3} (\frac{7}{4} t_{1}+4 t_{2}+\frac{7}{2} t_{3}+t_{j})}
|0100;010\rangle \nonumber \\
&&+ b |0100;100\rangle \, .
\end{eqnarray}

\subsection*{(v) recovery stage}
Generally, Alice and Bob may need the protocol to be repeated several times
until Alice registers only one click in the third stage.
It is easy to prove that after ${\cal N}$ repetitions of the protocol Bob's
system state is given by $a \theta |100\rangle_{B} +  b |010\rangle_{B}$
if Alice has detected one photon in the fourth stage and
$a \phi |010\rangle_{B} +  b |100\rangle_{B}$ if Alice has not registered
any detection in the fourth stage, where
$\theta=\epsilon \epsilon_{1} \alpha_{1}^{{\cal N}+1}
\alpha_{2}^{\cal N} \alpha_{3}^{-3}
\exp[\frac{i}{2} \delta_{3} (\frac{3}{2} t_{1}+t_{2}-13 t_{3}-2 t_{j})]
\mu_{0}^{{\cal N}_{0}} \mu_{2}^{{\cal N}_{2}}$,
$\phi=-\epsilon \epsilon_{1} \alpha_{1}^{{\cal N}+1}
\alpha_{2}^{{\cal N}+2} \alpha_{3}^{2}
\exp[\frac{i}{2} \delta_{3} (\frac{7}{2} t_{1}+8 t_{2}+7 t_{3}+2 t_{j})]
\mu_{0}^{{\cal N}_{0}} \mu_{2}^{{\cal N}_{2}}$,
$\mu_{0}=-\exp[i \frac{3}{2} \delta_{3} (t_{1}+t_{2})]$ and
 $\mu_{2}= \alpha_{3}^{-1} \exp[i \delta_{3} (t_{1}+3 t_{2}-3 t_{3})]$.
We denote by ${\cal N}_{0}$ and ${\cal N}_{2}$ numbers of repetitions caused
by zero and two-photon detections in the third stage.

In order to obtain the original state which Alice wanted to teleport,
the phase shift factor $\theta$ or $\phi$ has to be removed by Bob.
In case of no photon detection in the fourth stage Bob also has to
swap the amplitudes of his system states.
This is the objective of the fifth stage of the protocol.
During the stage both Alice's lasers are turned off.

In case of detection of one photon in the fourth stage Bob needs
three steps to remove the phase shift factor $\theta$.

\emph{(a)} First, Bob illuminates, using the $L_{B}$ laser,
his first atom for the time $t_{1}$ in order to perform
transformations~(\ref{eq:UD}) and~(\ref{eq:beta0}).

\emph{(b)} Next, he turns off the laser and waits for such time
$t_{\theta}$ that
$-\theta \alpha_{1}^{2} e^{i \Delta_{r} t_{\theta}}
e^{i 2 \delta_{3} t_{1}}=1$.

\emph{(c)} Finally, he again turns the $L_{B}$ laser on
illuminating his first atom for the time $t_{1}$.
In this way Bob performs transformations~(\ref{eq:UD_b}) and (\ref{eq:beta0}).

If no photon has been detected in the fourth stage Bob performs four
operations to remove the phase shift factor $\phi$ and to exchange the
amplitudes.

\emph{(a)} First, he turns the $L_{B}$ laser on and illuminates his
first atom for the time $t_{1}$. He performs
transformations~(\ref{eq:UD}) and~(\ref{eq:beta0}) in this step.

\emph{(b)} Next, he illuminates, using the $L_{B}$ laser, his second atom.
He turns the laser off after the time $t_{1}$ when
transformations~(\ref{eq:UD}) and~(\ref{eq:UD_b}) are done.

\emph{(c)} Next, he waits with lasers turned off for such time $t_{\phi}$
that $\phi e^{i \Delta_{r} t_{\phi}}=1$.

\emph{(d)} Finally, he illuminates his first atom using the $L_{B}$ laser.
He turns the laser off after the time $t_{1}$. In this way he performs
transformations~(\ref{eq:UD_b}) and (\ref{eq:beta0}).

After the last stage of the protocol Bob's system state is given by
$a |100\rangle_{B} +  b |010\rangle_{B}$.
\section{Numerical results}
\begin{figure}[htbp]
  \begin{center}
\includegraphics[width=8cm]{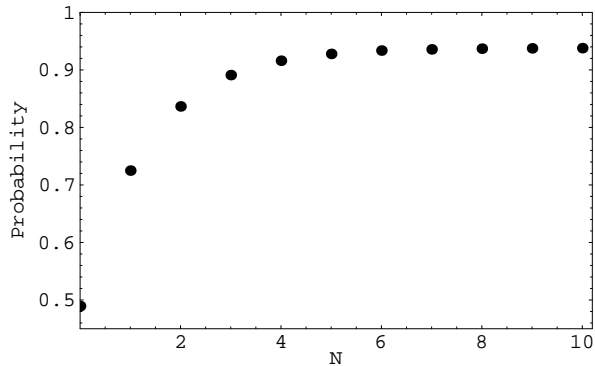}
    \caption{The average probability of success as a function of the number
    ${\cal N}$ of the repetitions of the first detection stage. }
    \label{fig:rysPN.eps}
  \end{center}
\end{figure}

In order to simplify the above considerations we have used some approximations
and therefore the fidelity of the teleported state and the probability of
the successful teleportation process have to be calculated numerically.
Both  quantities depend on the moduli of the amplitudes $a$ and $b$
of the initial state~\cite{bose,chimczak02:_effect}.
Therefore, we need to calculate average values of the fidelity and 
probability taken over all input states. We compute the averages using
the method of quantum trajectories~\cite{carmichaelksiazka,pleniotrajek}.
In order to take also into account such imperfections as spontaneous emission
from the excited states, we have performed the numerical calculations
with the full Hamiltonian~(\ref{eq:Hamiltonian0}).
We can get individual trajectory by generating a random initial state and
performing the whole teleportation protocol.
The initial state will be successfully teleported
when photon detections are only registered during the second step of
the preparation stage, the detection stage I or the second step of
the detection stage II. The state will be lost when either of the detectors
click during other steps of the protocol.
The initial state can be also destroyed by spontaneous atomic emission.
The trajectories in which the initial state is destroyed
are counted as the unsuccessful cases. The
average probability of a successful teleportation process is then given
by the ratio of the number of successful trajectories to the number of
the all trajectories. The average fidelity is taken over all successful
trajectories.
Before choosing numerical values for all parameters let us collect all
the aforementioned assumptions and rewrite them in a compact form
$({10}^{-1} \Delta \gg \Omega \gg \Omega^{\prime} \gg g;
\Delta^{\prime} \gg \gamma;
\delta_{5} \gg \kappa \gg \gamma \Omega^{2}/ \Delta^{2};
\Delta_{r}=\delta_{1})$. Now, it is easy to check that the parameter values
$(\Delta;\Omega;\Omega^{\prime};g;\gamma;\kappa)/2 \pi =
(2 \cdot {10}^{3};10;0.84;0.07;{10}^{-4};{10}^{-7})$ MHz satisfy the
conditions.
In order to make the average values reliable, we generate thirty thousand
trajectories. We have got the average fidelity about $F=0.984$ and the average
probability of success $P=0.94$.
These results show that the probability of
success is much higher than the successful teleportation probability in
other schemes~\cite{lee01:tel_ent,bose,chimczak02:_effect}. 
This is due to the fact that the initial state
is not lost in our scheme  when Alice's measurement is unsuccessful
contrary to the other schemes, when the initial state is lost and
the probability of success is equal or less than $0.5$. Owing to the
local redundant encoding 
technique used in our scheme the initial state is protected and therefore the
protocol can be repeated until only one photon is detected in the
detection stage.
Figure~\ref{fig:rysPN.eps} shows the probability to transmit the quantum state 
in the first try and in the subsequent repetitions.
As it is seen the probability that a single try will lead to the successful
transfer of the initial state is about $0.49$. Moreover, one can see
that the probability to achieve the successful teleportation process
after ${\cal N}$ repetitions saturates very quickly.
Therefore the protocol does not require great number of repetitions.

\begin{figure}[htbp]
  \begin{center}
\includegraphics[width=8cm]{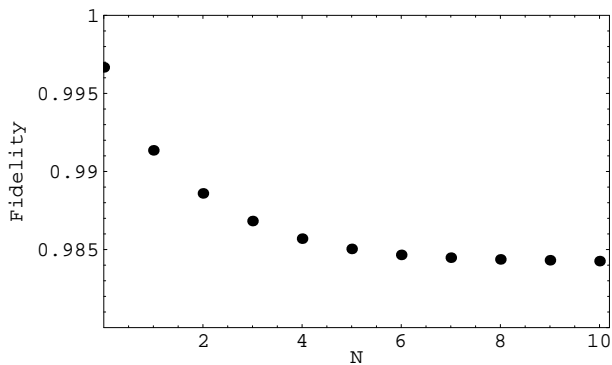}
    \caption{The average fidelity as a function of the number
    ${\cal N}$ of the repetitions of the first detection stage.}
    \label{fig:rysFN.eps}
  \end{center}
\end{figure}

As mentioned above, there are some imperfections in the encoding stage.
This is obvious that the imperfections decrease the average fidelity
of teleported state.
Also transformations recovering the original state can be done only
approximately.
In order to show the influence of the imperfections on the average
fidelity we plot the average fidelity as a function of the number of
the repetitions in Fig.~\ref{fig:rysFN.eps}.

One can see that the average fidelity decreases with
increasing ${\cal N}$. Thus, if higher fidelity is required, this can be
achieved by rejecting the cases with too high number of repetitions.
In order to show this improvement of average teleportation fidelity,
let us plot the average fidelity as a function the average probability.
\begin{figure}[htbp]
  \begin{center}
\includegraphics[width=8cm]{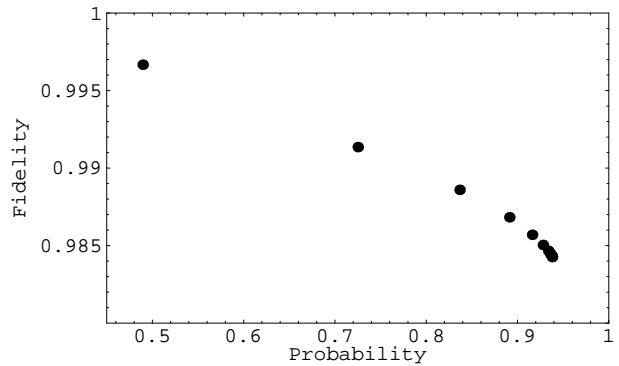}
    \caption{The average fidelity versus the average probability
    of success. The points are for ${\cal N}$=0,1,\dots}
    \label{fig:rysFP.eps}
  \end{center}
\end{figure}
As it is evident from Fig.~\ref{fig:rysFP.eps} the increase of
average fidelity can be achieved by accepting lower success rate.
Moreover, the increase of the
average fidelity and the decrease of the average probability
is higher for small number of cases counted as successful.
When the repetition number limiting
the successful cases is high the points become indistinguishable.
Therefore the teleportation scheme will work properly even when we
set maximal number of repetition to six as clearly illustrated
in Fig.~\ref{fig:rysFP.eps}.

\section{Conclusions}
In this paper we have presented a scheme performing quantum teleportation
of atomic entangled states via cavity decay.
The distinguishing feature of our protocol is using the
local redundant encoding technique. We have shown
feasibility of the technique in detail for atoms trapped in a cavity
and manipulated by laser fields. Since the technique codes the initial state
in the way that the state is secure during the detection stage,
the encoding procedure and the detection stage can be repeated until
the teleportation is successful. The numerical calculations show that
the average probability of success of the protocol is about $0.94$
while the average probability of successful teleportation
without the insurance does not exceed $0.5$. Moreover, we have shown
that not more than six repetitions are enough to obtain high average values of the
probability and the fidelity of the teleportation.
We have also shown that although the average fidelity is as high as $0.984$,
one can still increase it by rejecting the cases with too many repetitions
and accepting lower success rate.
In addition, we have shown how to manipulate states of many atoms trapped
in a cavity using two lasers. We believe that the analytical results presented in
Sec. III can be helpful for description of various atomic systems in optical cavities.

\end{document}